\documentclass[10pt]{article}

\usepackage[utf8]{inputenc}   
\usepackage{newtxtext,newtxmath}
\usepackage{geometry}      
\usepackage{multicol}
\usepackage{graphicx}      
\usepackage{array}
\usepackage{tabularx}
\usepackage{multirow}
\usepackage{caption}

\usepackage{scalerel}
\usepackage{tikz}

\usepackage[colorlinks=true, linkcolor=black, urlcolor=blue, citecolor=blue]{hyperref}
\usepackage{enumitem}
\usepackage{cite}

\begin{document}

\begin{center}

\noindent \begin{minipage}[c]{0.85\textwidth}
\vspace{25mm}
\Large\textbf{Atomically precise mechanosynthesis of carbon structures on hydrogenated Si(100) by inverted-mode STM}
\vspace{1em} 
\end{minipage}
\noindent \begin{minipage}[c]{0.86\textwidth}
\vspace{2mm}
\setcounter{page}{1}
\makeatletter
\centering 
\small\textbf{\noindent \raggedright Megan Cowie*, Chris Deimert, Ryan Groome, Alex Inayeh, Robert J. Kirby, Cameron J. Mackie, Jonathan Myall, Sam Rohe, Luis Sandoval, Khalil Sayed-Akhmad, Bheeshmon Thanabalasingam, Reid Wotton, Rafik Addou, Aly Asani, Brandon Blue, Adam Bottomley, Kareem A. Clarcia, Tyler Enright, James Zhangming Fan, Robert A. Freitas Jr., Alan T.K. Godfrey, Si Yue Guo, Aru Hill, Taleana Huff, Mark Jobes, Hadiya Ma, Adam C. Maahs, Oliver MacLean, Steven M. Maley, Michael Marshall, Terry McCallum, Ralph C. Merkle, Mathieu Morin, Ryan Plumadore, Henry Rodriguez, Marc Savoie, Benjamin Scheffel, Janice L. Wong, Damian G. Allis, Jeremy Barton, Michael Drew, Matthew R. Kennedy, Tait Takatani, Marco Taucer, Dusan Vobornik, Ryan Yamachika, Mathieu Durand}

\vspace{5mm}

\normalsize\textit{CBN Nano Technologies, Inc. (CBNNT); Ottawa, K1Y 4W5, Canada}

\vspace{2mm}
\today

\end{minipage}

\vspace{5mm} 

\noindent \begin{minipage}[c]{0.85\textwidth}
\normalsize\noindent The ability to build atomically precise structures on surfaces with complete control over both atomic placement and chemical bonding remains a central challenge in nanoscale fabrication. Here, we demonstrate simultaneous spatial and chemical control over the mechanosynthetic fabrication of carbon structures. Using inverted-mode STM, C$_2$ units are donated from surface-deposited molecules to pre-patterned reactive sites on a hydrogen-passivated Si(100) surface. We demonstrate single-site C$_2$ donation, spatially patterned multi-site C$_2$ donation, and the stepwise assembly of polyyne structures through successive C-C bond formation. Together, these results establish controlled mechanosynthetic donation as a foundational capability for programmable atomically precise fabrication.
\end{minipage}
\end{center}

\vspace{12mm}

\noindent Achieving control over matter at the level of individual atoms is a defining vision of nanotechnology, underpinning the development of molecular-scale electronics, artificial lattices, and semiconductor quantum devices, with the potential for order of magnitude improvements in speed and energy efficiency over current devices\cite{Feynman1960, Khajetoorians2019, Chatterjee2021, Schofield2025}. In these emerging technologies, function derives from atomic composition, bonding, and spatial organization, and realizing full performance potential requires deterministic structural control at the atomic scale. Atomically precise fabrication (APF) seeks to realize this level of control through atom-by-atom operations within a fixed reference frame, typically on a surface. Unlike ensemble‑driven synthesis or top‑down fabrication, APF enables the bottom-up construction of positionally defined structures that can be repeatedly re‑addressed for interrogation or modification, linking atomic structure to technological function. 
\vspace{1em} 

\noindent Scanning probe microscopy (SPM) is an established platform for pursuing APF, having enabled manipulations of individual atoms and molecules for more than three decades\cite{Eigler1990, Crommie1993, Beton1995, Lyding1996, Jung1996, Oyabu2005}. SPM also enables synthesis through tip‑induced reactions between individual molecules, creating extended covalently bonded carbon structures\cite{Hla2000, Okawa2012, Pavlicek2018, Kaiser2019, Kawai2020, Zhong2021, Albrecht2022}. However, existing approaches are largely confined to surface‑bound species, with resulting structures extending along the surface plane. Control beyond the surface plane is essential to building many realistic devices. While limited three-dimensional control by vertical manipulations, including lifting and repositioning of adsorbates and transfers of individual atoms, has been demonstrated, reliable SPM‑enabled synthesis leveraging out‑of‑plane bonding to build extended structures remains a key challenge\cite{Bartels1997, Dujardin1998, Oyabu2003, Pump2008, Gross2009, Sugimoto2008, Huff2017, Cai2025}.
\vspace{1em} 

\noindent Positionally controlled mechanosynthesis offers a route to this goal: it is a mode of synthesis in which chemical species are positioned and oriented in three dimensions to direct bond formation without reliance on thermal, optical, or electronic excitation\cite{Drexler1992, Freitas2011}. Individual mechanosynthetic operations have recently been demonstrated, but not yet in the controlled assembly of extended covalent structures\cite{Bothra2023, Barrera2025, Blue2026}. Achieving controlled out-of-plane assembly requires (i) reliable placement of reactive fragments at predetermined sites and (ii) iterative chemical operations on the resulting structures. Here, we demonstrate both capabilities by deterministically placing C$_2$ units on a hydrogenated silicon surface and extending these structures through successive C–C bond formation by vertical transfer of an additional C$_2$.
\vspace{1em}

\vspace{1em} 
\noindent\textbf{Mechanosynthetic C$_2$ donation}
\vspace{1em} 

\noindent Mechanosynthesis was performed using inverted‑mode scanning tunneling microscopy (IM‑STM) operated at 4~K, following \cite{Barrera2025}. In this approach, sparsely deposited tall molecules on the sample surface act as probes that image the apex of a large, flat silicon probe chip (SPC) (Fig. 1A). The molecules can transfer fragments to and from the SPC apex, which serves as a build site\cite{Barrera2025, Blue2026}. The dual functionality of these molecules – as imaging probes and chemical reactants – is the basis for referring to them as ``molecular tools". The molecular tool used here, EAOGe-C$_2$I, consists of a Ge-substituted adamantane with a C$_2$ functional group, three OH-terminated legs, and an iodine capping group\cite{McCallum2026}. The OH-terminated legs anchor the molecules to the Si sample, orienting a subset upright with the Ge–C$_2$–I moiety perpendicular to the surface (Fig. 1A); this configuration appears tallest in IM-STM\cite{Barrera2025, Kirby2026}. 
\vspace{1em} 

\begin{figure*}[h!]
    \centering
    \includegraphics[width=\textwidth]{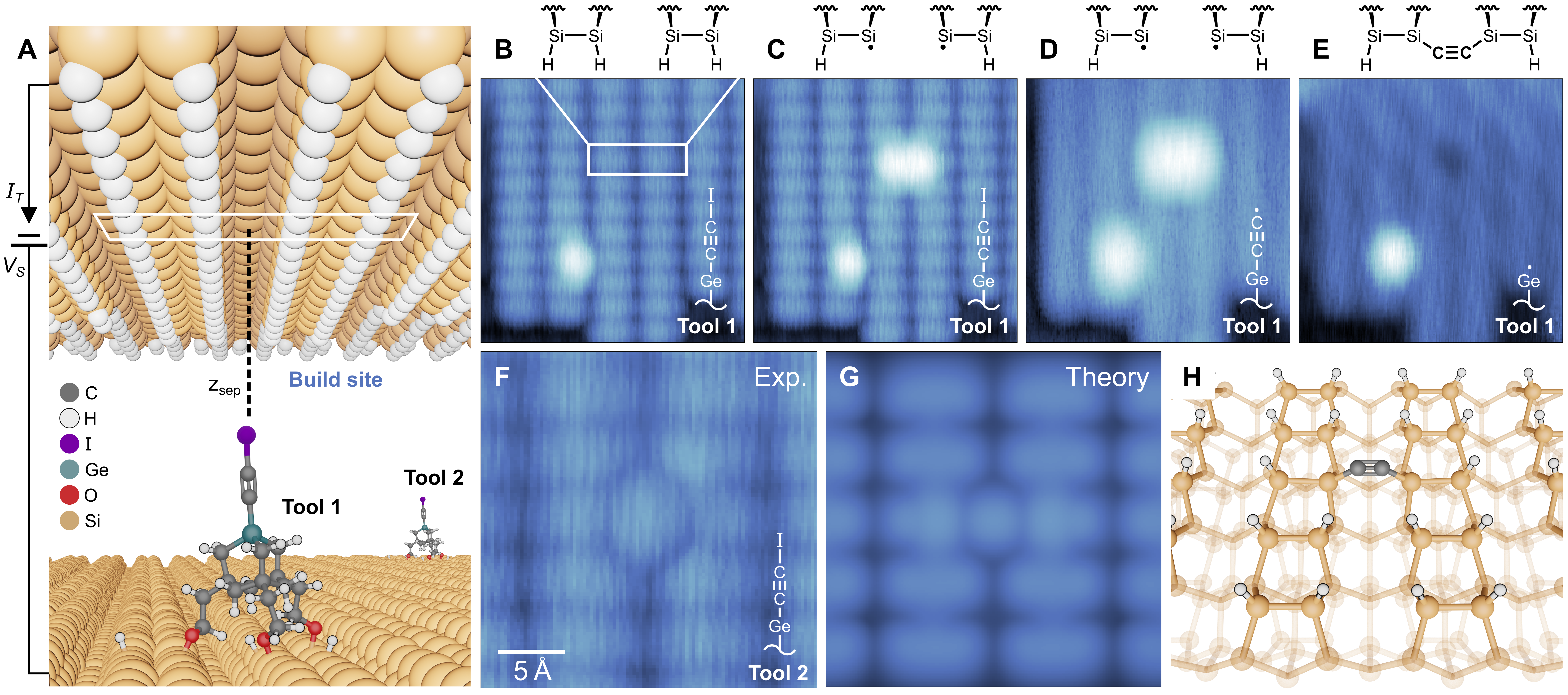}
    \caption{\textbf{Mechanosynthetic C$_2$ donation.} \textbf{(A)} Schematic of the inverted-mode STM setup. EAOGe-C$_2$I molecules are deposited on flat Si(100), and an H-passivated Si(100) silicon probe chip (SPC) with a flat, crystalline apex is positioned above the surface. The molecules function both as imaging probes, where an applied bias (\textit{V\textsubscript{S}}) drives a tunneling current (\textit{I\textsubscript{T}}) through the molecule, and as reagents capable of chemically reacting with the build site. \textbf{(B-E)} Mechanosynthetic sequence showing the evolution of the build site as it is imaged with and modified by a molecule (Tool 1). As the build sequence proceeds, both the tool termination (inset) and atomic composition in the target area (white rectangle) change. Starting from a bare build site (B), two Si dangling bonds (DBs) are patterned by bias pulsing in an inter-row (IR) configuration (C), followed by molecule de-iodination (D), and transfer of a C$_2$ unit to the DB pattern (E). \textbf{(F)} Small-area scan of the build site following the C$_2$ transfer, centered on the target area and imaged with a new, intact tool (Tool 2). (G, H) Simulated STM image and geometry of C$_2$ in the IR configuration (IR-C$_2$), reproducing the experimental image shown in (F).}
\end{figure*}

\noindent Figure 1B-E shows a mechanosynthetic C$_2$-donation sequence from a EAOGe-C$_2$I molecule. Changes in the termination of Tool~1 produce corresponding changes in the IM-STM contrast, analogous to the dependence of imaging contrast on probe termination in conventional STM. This evolving ``imaging modality" is therefore a direct indicator of the evolving Tool 1 termination\cite{Barrera2025}. These imaging modalities are reproducible across C$_2$ donation sequences (see Supplementary Materials for examples). A single silicon dangling bond (DB) is present near the target area; this bright feature remains chemically unchanged and serves as a fixed reference for the evolving composition of the target area.
\vspace{1em} 

\noindent The build sequence begins on an H-terminated, 2\texttimes1-reconstructed Si(100) (H:Si) surface (Fig. 1B). A DB pair is created in an inter-row (IR) configuration (spanning a trough of the reconstructed H:Si surface) (Fig. 1C). DB formation follows the established electron-induced desorption mechanism, with the molecular tool simply taking the place of a conventional STM tip (see Supplementary Materials)\cite{Shen1997}. DBs can also be patterned by mechanosynthetic H abstraction, as demonstrated in \cite{Barrera2025}. With the IR-DB pair in place, Tool 1 is de-iodinated away from the build site by increasing the bias, thus exposing the reactive C$_2$ radical (•)\cite{Barrera2025}. De-iodination produces a change in imaging modality (Fig.~1D): EAOGe-C$_2$I resolves individual H:Si dimers, but EAOGe-C$_2$• resolves only the dimer rows. 
\vspace{1em} 

\noindent At this stage, the system is prepared for mechanosynthesis. The de-iodinated Tool 1 is aligned under the centre of the IR-DB pair. The bias is set to 0 V, eliminating tunneling current, and the STM z-controller is disengaged. Mechanosynthesis is then carried out using iterative depth sampling with repeated approach-retraction cycles from a large initial separation. The maximum approach depth is increased in 50 pm increments (less than half the $\sim$120~pm C$_2$-dimer length), and a constant-current IM-STM image is acquired after each increment. This is repeated until a change in imaging modality and apparent STM height is observed, indicating a change in the termination of Tool~1 (Fig. 1E). At this point, a dark feature appears at the former IR-DB site; this can be contrasted to the nearby unmodified DB, which remains bright.
\vspace{1em} 

\noindent To characterize the transferred product, the SPC is repositioned over a nearby intact molecule (``Tool 2''), and a close-up IM-STM image of the build site is acquired (Fig 1F). The image reveals a new feature in the target area, which is assigned to C$_2$ in an IR configuration (IR-C$_2$). This assignment is supported by multiple lines of evidence. First, the imaging modalities observed during the build sequence are distinct from those associated with other reaction outcomes, such as hydrogen abstraction from nearby sites, which produces a qualitatively different IM\nobreakdash-STM appearance (see \cite{Barrera2025} and Supplementary Materials). These modalities are highly reproducible across build sequences (see Supplementary Materials). Second, a filled-state STM simulation of IR-C$_2$ (Fig. 1G) closely matches the experimental image, including a hexagonal feature with twofold mirror symmetry spanning a surface trough and extending onto adjacent dimers, with no significant height contrast relative to the surrounding H:Si. Third, the feature is inconsistent with known surface defects, which are not trough-centred, and with symmetries of alternative C$_2$ binding geometries on H:Si (see Supplementary Materials)\cite{Croshaw2020}. Fourth, the same IR-C$_2$ configuration has been produced and assigned on un-passivated Si(100) via de-hydrogenation of adsorbed acetylene and ethylene\cite{MacLean2026}. Finally, density functional simulations predict formation of IR-C$_2$ under the experimental conditions used here (shown below). Together, the evolution of imaging modality, agreement between experiment and STM simulation, exclusion of alternative structures, independent synthesis, and the predicted formation pathway provide strong evidence for the assigned composition and binding.
\vspace{1em} 

\vspace{2em} 
\noindent\textbf{Stepwise C$_2$ patterning}
\vspace{1em} 

\noindent This method of build site pre-patterning combined with mechanosynthetic donation enables site-specific placement of C$_2$ on silicon. Importantly, this capability can be extended through repetition: following IR-C$_2$ donation, the SPC is repositioned over a new, intact molecular tool to initiate another IR-C$_2$ donation sequence. Iterating this sequence achieves atomically precise patterning of the silicon surface with carbon (Fig. 2), establishing stepwise mechanosynthetic donation with atomic‑scale control.
\vspace{1em} 

\begin{figure*}[h!]
    \centering
    \includegraphics[width=0.31\textwidth]{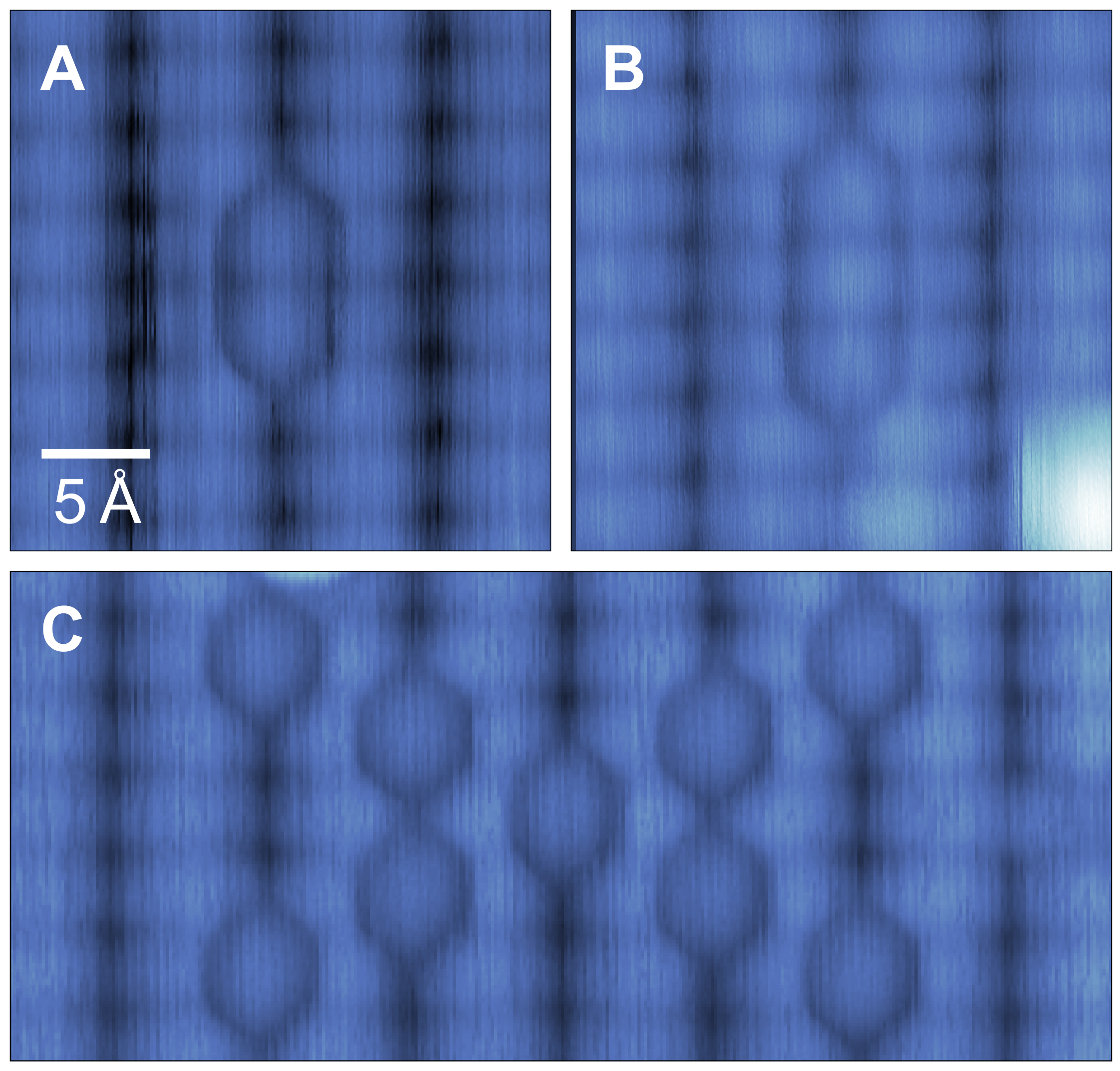}
    \caption{\textbf{IR-C$_2$ patterning.} IM-STM images showing \textbf{(A)} two and \textbf{(B)} three IR-C$_2$ units patterned side-by-side along a dimer row, and \textbf{(C)} nine C$_2$ units, comprising 18 carbon atoms, patterned in an `X' shape (\textit{V\textsubscript{S}}~=~+3.2 V, \textit{I\textsubscript{T}}~=~10 pA, imaged with a EAOGe-C$_2$I molecule).}
\end{figure*}

\noindent Once patterned, IR‑C$_2$ units remain intact for days to weeks of 4 K IM‑STM operation, with no observed changes over the measurement period. IR-C$_2$ withstands a wide range of imaging conditions, including sample biases from approximately -3 to +4 V, tunneling currents from 10-100 pA, and imaging with molecular tools of varied terminations (EAOGe-C$_2$I, EAOGe-C$_2$•, EAOGe•, and EAOGe-C$_2$H). This stability likely arises from the absence of reactive partners and the lack of energetically accessible rearrangement pathways. The calculated geometry of IR‑C$_2$ (Fig. 1H) is predicted to be less strained than alternative adsorption configurations on Si(100) (see Supplementary Materials for geometries of comparison systems)\cite{MacLean2026}.
\vspace{1em}

\vspace{1em} 
\newpage
\noindent\textbf{C$_2$ donation mechanism}
\vspace{1em} 

\noindent Figure 3 shows a proposed C$_2$ donation mechanism, based on a quantum mechanics / molecular mechanics (QM/MM) model, connecting panels (D) and (E) of Fig. 1\cite{Senn2009, Svensson1996}. At the start of mechanosynthesis, a de\nobreakdash-iodinated molecular tool (EAOGe-C$_2$•) is positioned with the distal C atom centred under an IR-DB pair (Fig.~3A). The separation of the two Si surfaces (z) decreases as the molecular tool approaches the build site. At a critical separation (z=z$_0$), the first C–Si bond forms (Fig. 3B), producing a sharp drop in potential energy. Upon retraction (z increases), the strained molecular junction cleaves at the Ge–C bond (Fig. 3C). This bond-breaking event results in the transfer of C$_2$ from the molecular tool to the build site, producing an upright, surface-bound pendent C$_2$• intermediate which is transiently stabilized by a coordinative radical interaction with the remaining EAOGe• molecular tool. With further retraction, this interaction weakens until the relaxation of the pendent C$_2$• into the IR configuration becomes barrierless (see Supplementary Materials for supporting calculations), resulting in the formation of IR-C$_2$ (Fig. 3D). The key takeaway is that the transfer of C$_2$ from the molecular tool to the build site proceeds along a net downhill energy landscape, and all barriers in the reaction pathway are overcome by the mechanical work performed by varying z.

\begin{figure*}[h!]
    \centering
    \includegraphics[width=0.658\textwidth]{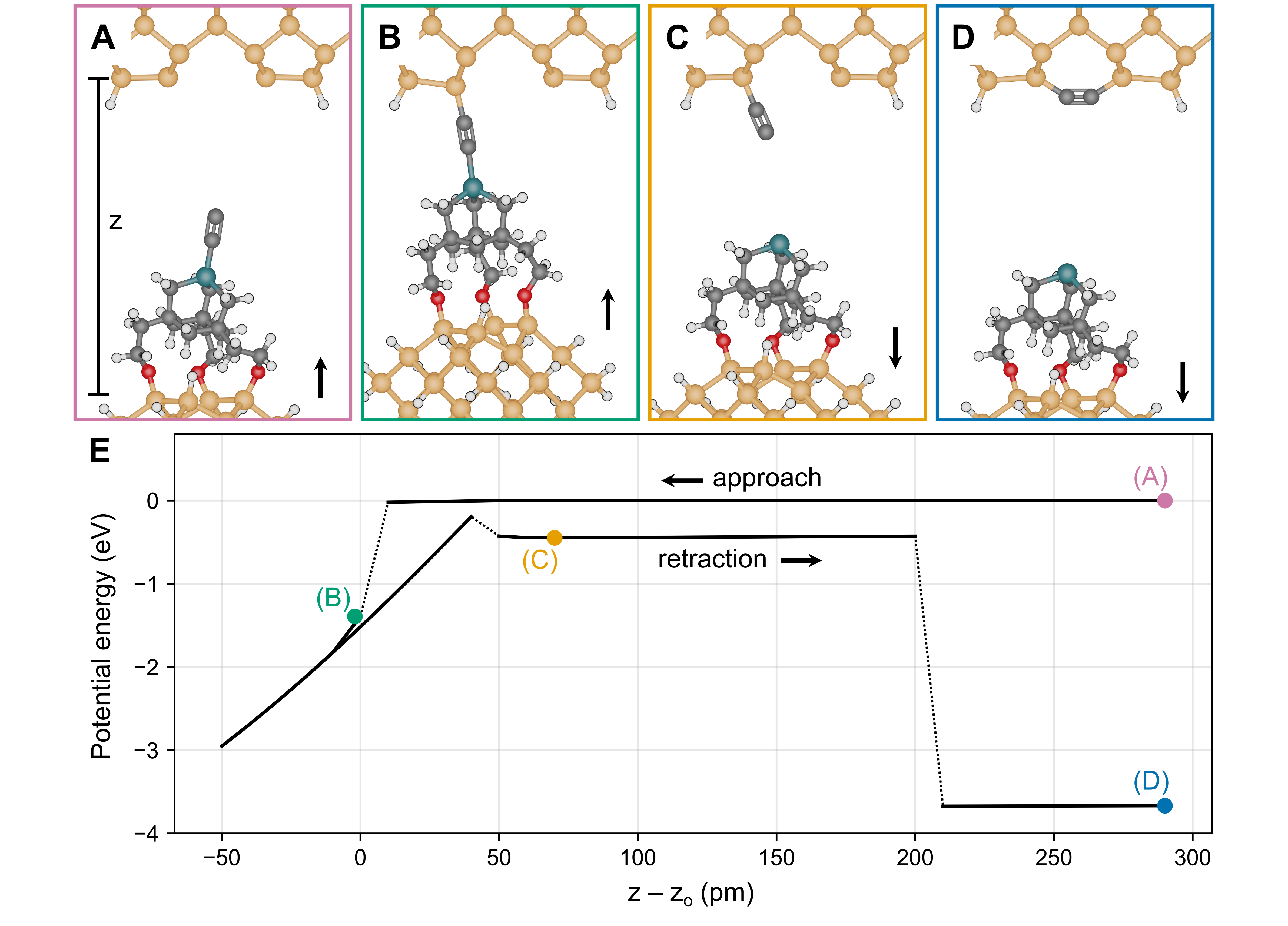}
    \caption{\textbf{IR-C$_2$ formation mechanism.} QM/MM xTB(GFN0)/DFT ($\omega$B97X-D3) simulation illustrating the proposed mechanosynthetic IR-C$_2$ donation mechanism. \textbf{(A-D)} Atomic configurations of the molecular tool (bottom) and the target area of the SPC (top) at key stages of mechanosynthesis: positioning of the de-iodinated molecular tool (EAOGe-C$_2$•) over an inter-row dangling bond (IR-DB) pair (A), formation of the first C–Si bond at z=z$_0$ during approach (B). C$_2$ donation following Ge–C bond cleavage upon retraction (C), and relaxation into the IR-C$_2$ configuration (D). \textbf{(E)} Energy profile as function of z-z$_0$, with points corresponding to configurations (A-D) indicated. This arrangement represents only one of several possible low-energy leg binding configurations and is shown here as a representative example. An animation of this process is provided in the Supplementary Materials.}
\end{figure*}

\noindent The experimental depth-sampling protocol, in which progressively smaller z values are sampled in 50 pm increments, limits the minimum sampled separation such that z-z$_0$ $>$ –50 pm. At this separation, the Ge–C$_2$–Si moiety (Fig. 3B) experiences only a temporary, reversible compression relative to the point of initial Si–C bond formation (z-z$_0$ = 0); otherwise, the proposed donation mechanism is unchanged. At larger approach depths, IR-C$_2$ formation still occurs but can proceed via an alternative pathway that involves formation of the second Si–C bond while the carbon is still engaged in Ge–C bonding (see Supplementary Materials for supporting calculations). 
\vspace{1em}

\vspace{1em} 
\noindent\textbf{Extension of carbon structures by C–C bond formation}
\vspace{1em} 

\noindent This mechanosynthesis strategy is not limited to placement of isolated C$_2$ units on the Si(100) surface. Once transferred, surface‑bound IR‑C$_2$ structures can undergo successive reactions. Figure 4A shows an STM feature resulting from a mechanosynthetic build sequence starting from IR-C$_2$. The procedure is similar to that of IR‑C$_2$ formation (Fig. 1), except that the de‑iodinated tool (EAOGe‑C$_2$•) is positioned under an IR‑C$_2$ unit rather than an IR-DB pair. Throughout the build sequence, the imaging modalities associated with the evolving molecular tool termination are consistent with those observed during IR‑C$_2$ donation (see Supplementary Materials for examples).  
\vspace{1em} 

\begin{figure*}[h!]
    \centering
    \includegraphics[width=\textwidth]{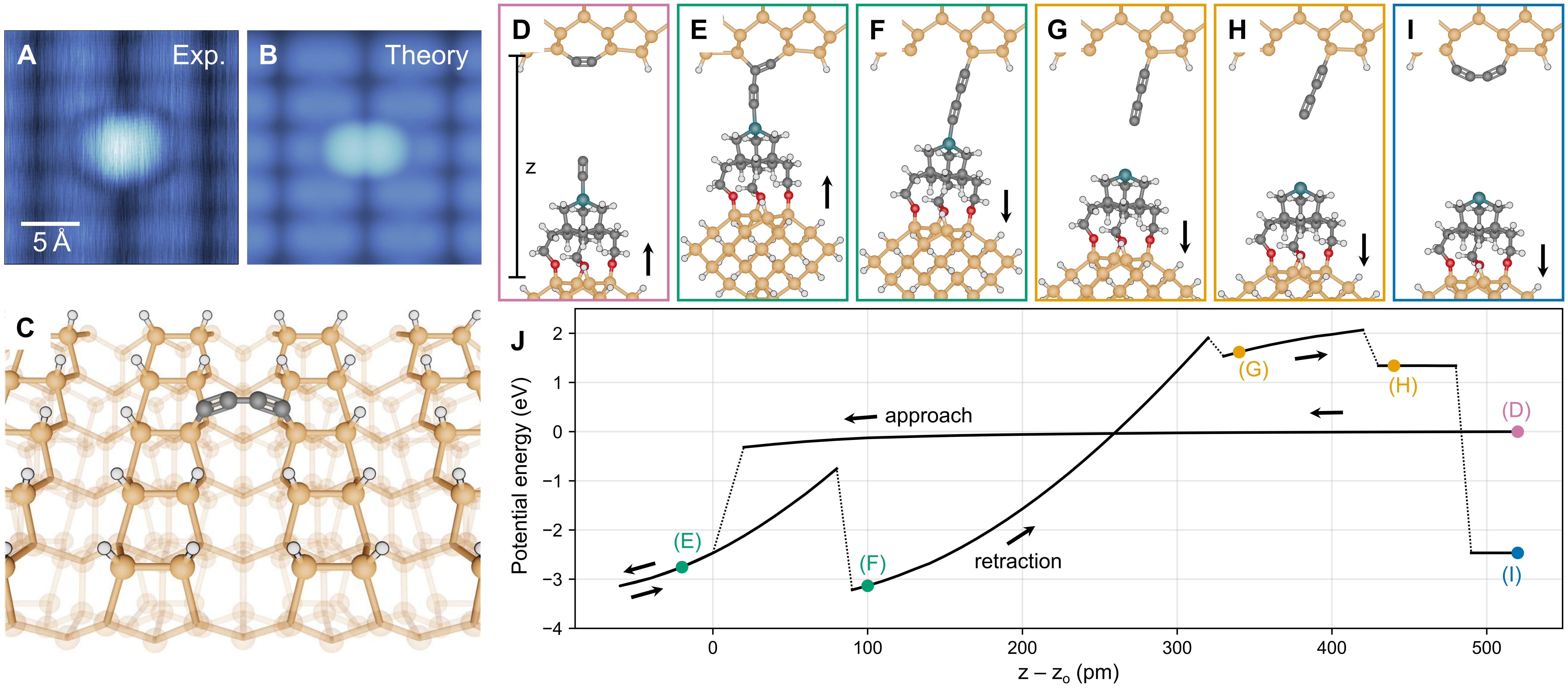}
    \caption{\textbf{IR-C$_4$ formation.} \textbf{(A)} Experimental feature observed following C$_2$ donation to an existing IR-C$_2$ structure. \textbf{(B,C)} Simulated STM image and geometry of IR-C$_4$. \textbf{(D-J)} QM/MM xTB(GFN0)/DFT ($\omega$B97X-D3) simulation illustrating the proposed mechanosynthetic IR-C$_4$ formation mechanism, via extension of a surface-bound IR-C$_2$. Atomic configurations of the molecular tool (bottom) and the target area of the SPC (top) at key stages of mechanosynthesis are shown: positioning of the de-iodinated molecular tool (EAOGe-C$_2$•) under an IR-C$_2$ structure (D), formation of a four-carbon chain via C–C bonding at z=z$_0$ (E,F), C$_2$ chain extension via C$_2$ donation following Ge–C bond cleavage upon retraction (G,H), and relaxation into the IR-C$_4$ configuration (I). (J) Potential energy profile as function of z-z$_0$, with the points corresponding to configurations (D-I) indicated. An animation of this process is provided in the Supplementary Materials.}
\end{figure*}

\noindent The resulting feature is assigned to a four‑carbon chain in an IR configuration (IR‑C$_4$). This assignment is supported by multiple lines of evidence. First, imaging modalities are consistent with C$_2$ donation. Second, experimental STM images agree with simulations (Fig. 4A,B), both showing a symmetric feature with pronounced height contrast relative to the surrounding H:Si. A nodal feature present in the simulated image is not observed experimentally; this is attributed to limitations in spatial resolution (see Supplementary Materials). Third, IR-C$_4$ is the lowest energy configuration among alternative geometries by more than 2 eV (see Supplementary Materials). Fourth, trajectory calculations predict its formation. Notably, IR‑C$_4$ has not been reported previously, which highlights that mechanosynthesis enables the creation of novel surface-bound structures in addition to placement at defined sites.
\vspace{1em} 

\noindent Figure 4D-J shows a proposed QM/MM mechanism for mechanosynthetic extension of IR‑C$_2$ to IR‑C$_4$. At the start of mechanosynthesis, the de-iodinated tool (EAOGe-C$_2$•) is positioned centrally under an IR-C$_2$ unit (Fig.~4D). Upon approach, the molecular tool undergoes radical addition to the IR-C$_2$, resulting in the formation of an intermediate vinyl radical (Fig. 4E).  During retraction, the intermediate vinyl radical partially detaches from the underlying Si, rearranging into a four-carbon diyne intermediate bound to both the Si surface and the molecular tool (Fig. 4F). Calculations suggest that continued retraction leads to rupture of the Ge–C bond, producing a transient, coordinatively stabilized pendent C$_4$• structure (Fig. 4G), which transitions to an alternative pendent C$_4$• configuration as the separation increases (Fig. 4H). At a sufficiently large separation, the pendent C$_4$• intermediate relaxes into an IR configuration, forming IR-C$_4$ (Fig.~4I). 
\vspace{1em}

\vspace{2em} 
\noindent\textbf{Reproducible mechanosynthesis of extended carbon structures}
\vspace{1em} 

\noindent The IR-C$_4$ structure, like IR-C$_2$, is robust to a range of imaging conditions with variable tool terminations, with no observed changes over the measurement timescale (days). This stability enables its use as a building unit for atomically precise patterning of the silicon surface. Figure 5 shows a build sequence producing two adjacent IR‑C$_4$ units along an H:Si dimer row. Four successive C$_2$ transfers generate an eight‑carbon pattern with site‑specific placement. The sequence proceeds through IR‑C$_2$, 2IR‑C$_2$, IR‑C$_2$/C$_4$, and finally 2IR‑C$_4$. Each step in the build sequence has been replicated between tens and more than 100 times, demonstrating reproducibility (see Supplementary Materials for examples). Figs. 5I–L summarize outcomes for each step, grouped into successful target formation, undesired C$_2$ products, H‑abstraction from the surrounding H‑passivated Si surface, and other (generally non-reproducible) outcomes.

\begin{figure*}[h!]
    \centering
    \includegraphics[width=0.658\textwidth]{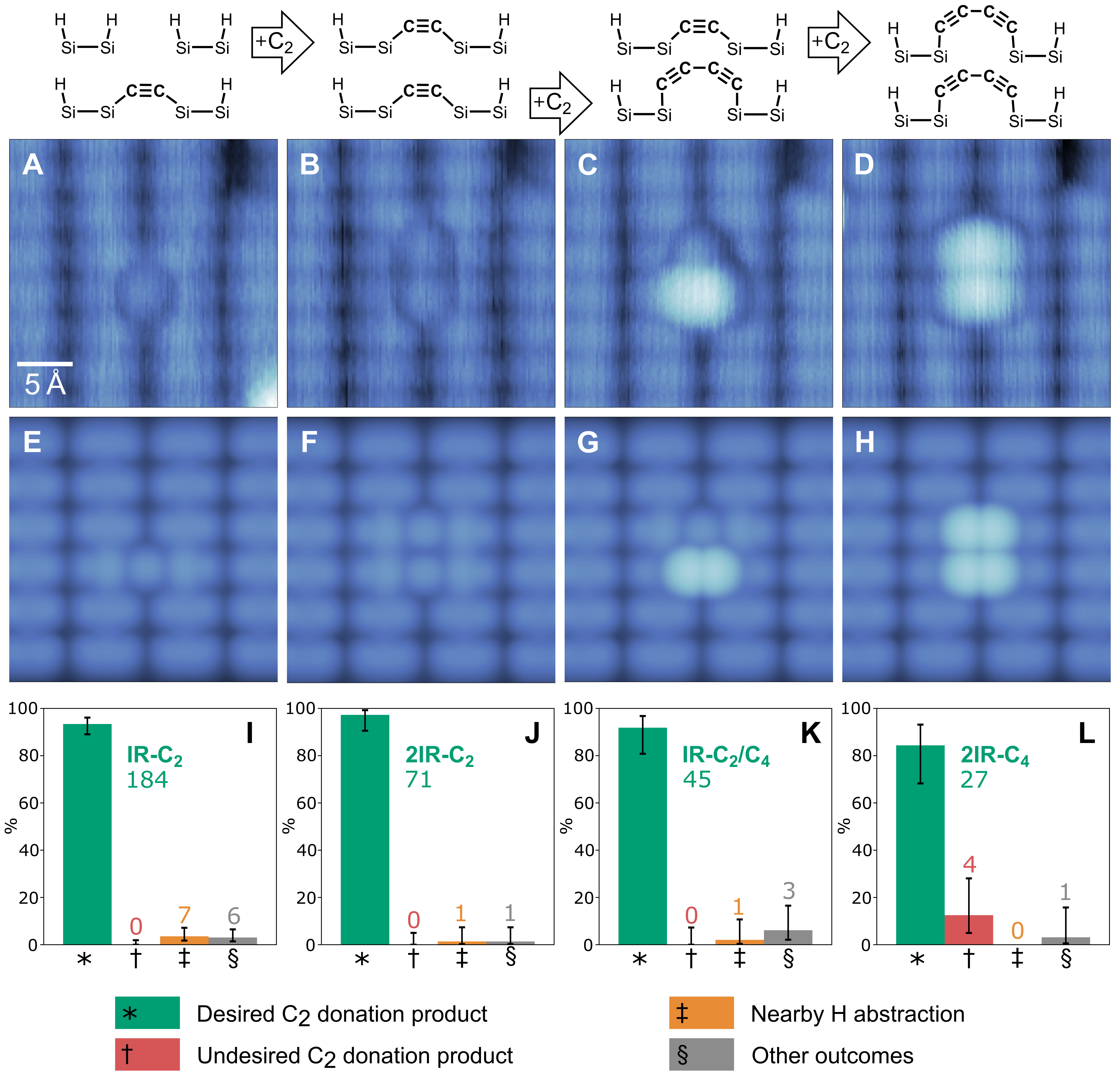}
    \caption{\textbf{Mechanosynthesis of complex carbon patterns.} \textbf{(A-D)} Experimental STM images showing a representative build sequence. Starting from an isolated IR‑C$_2$ (A), a second IR‑C$_2$ is positioned adjacent along a dimer row (2IR‑C$_2$, B). Sequential C–C bond formation then extends one IR‑C$_2$ to form IR‑C$_4$ adjacent to IR‑C$_2$ (IR‑C$_2$/C$_4$, C), followed by extension of the remaining IR‑C$_2$ to yield side‑by‑side IR‑C$_4$ units (2IR‑C$_4$, D). The atomic configuration of the target build region is shown above each image. \textbf{(E-H)} Simulated STM images of the corresponding structures. \textbf{(I-L)} Outcome statistics for each mechanosynthetic interaction, starting from the required precursor (IR-DB pair, IR-C$_2$ adjacent to an IR-DB pair, 2IR-C$_2$, and IR-C$_2$/C$_4$, respectively). The plots report percentages with annotated counts; error bars indicate 95\% confidence intervals computed using the Wilson score.}
\end{figure*}

\newpage
\noindent\textbf{Positional and chemical control of mechanosynthetic donation}
\vspace{1em} 

\noindent The build sequence in Fig. 5 demonstrates positional and chemical control over mechanosynthetic C$_2$ donation, with high yields: 93\% for IR‑C$_2$ (184/197, 95\% confidence interval 89-96\%), 97\% for 2IR‑C$_2$ (71/73, 95\% confidence interval 91-99\%), 92\% for IR‑C$_2$/C$_4$ (45/49, 95\% confidence interval 81-97\%), and 84\% for 2IR‑C$_4$ (27/32, 95\% confidence interval 68-93\%). These targeted outcomes contrast the broad distribution of C$_2$ donation products observed on un-passivated, un-patterned Si(100), highlighting a key aspect of reaction sequence design: achieving control in mechanosynthesis depends as much on the preparation of the build site – in this case, DB patterning of H:Si – as it does on the choice of molecular tool and transferred group\cite{Blue2026}. 
\vspace{1em} 

\noindent Successful mechanosynthetic C$_2$ donation can be decomposed into the following mechanistic steps: (i) formation of the desired initial bond (Si–C for IR-C$_2$ and C–C for IR-C$_4$), (ii) cleavage of the Ge–C bond, and (iii) relaxation of the resulting proposed pendent intermediate structure (C$_2$• or C$_4$•) into the desired IR configuration. The high yields of IR-C$_2$ and IR-C$_4$ indicate that each of these steps occurs with high fidelity. Next, we examine each step in turn and identify how off‑target outcomes arise when processes deviate from the intended pathway. 
\vspace{1em} 

\noindent For the initial approach with EAOGe-C$_2$•, the intended process is the formation of a Si–C or C–C bond (i), as shown in Figs. 3A,B and 4D,E, respectively. The only predicted alternative bonding process at this stage is formation of a C–H bond between the EAOGe-C$_2$• tool and a hydrogen atom on a nearby passivated Si site, which results in H-abstraction via the mechanism reported in \cite{Barrera2025}. This H-abstraction pathway was observed in nine out of the 351 total interactions shown in Fig. 5, making it one of the most common off-target outcomes (3\%, 95\% confidence interval, 1-5\%). Several factors could contribute to these undesired H-abstraction events: lateral flexibility of the molecular tool, hydrogen tunneling to the EAOGe-C$_2$• molecule, or limited positional control during the approach (see Supplementary Information for lateral targeting tolerance). 
\vspace{1em} 

\noindent The next mechanistic step required for C$_2$ donation is cleavage of the Ge–C bond (ii), see Figs. 3B,C and 4F,G. As the tool is retracted, three covalent bonds are subject to competing tensile strain: the Ge–C bond, the Si–C bond, and the three Si–Si bonds anchoring the C-bound Si atom to the lattice. Empirically, the binding forces within the H:Si lattice are sufficiently strong that the Si–Si bonds do not break during tool retraction: abstraction of a Si atom from the build site is almost never observed, and is believed to account for only one out of the 351 total interactions in Fig.~5 (0.3\%, 95\% confidence interval, 0.05-2\%). The Ge–C and Si–C bonds are predicted to be very close in energy based on molecular proxy calculations (5.200 vs. 5.139 eV at the $\omega$B97X-D/Def2-TZVPP level). Despite this predicted near-degeneracy, the empirical frequency of C$_2$ donation shows that Ge–C bond cleavage does occur preferentially during tool retraction. 
\vspace{1em} 

\noindent The third and final mechanistic step is relaxation of the intermediate pendent C$_2$• or C$_4$• structure into the IR configuration (iii), see Figs. 3C,D and H,I. The orientation of the pendent relative to the surface is governed by two factors: the preferred sp$^3$ hybridization geometry of the anchoring Si atom and the lateral position of the post-donation tool (EAOGe•), which maintains a stabilizing coordinative radical interaction with the pendent structure (see Supplementary Materials for supporting calculation). Together, these constraints tilt the pendent intermediate over the trough toward the reactive site of the remaining DB, rather than neighbouring dimers, where radical coupling then drives relaxation to the IR configuration. Other possible relaxations of the pendent C$_2$• or C$_4$• structures, which could abstract a nearby H atom from the surface, resulting in a pendent C$_2$H or C$_4$H, have not been observed.
\vspace{1em} 

\noindent A single reproducible off‑target C$_2$ donation product is observed during attempts to form 2IR-C$_4$ (see Supplementary Materials). This off-target structure likely arises from a failure mode in either the initial bond‑formation step (i) or the final relaxation step (iii). Two possible failure pathways are proposed. In the first, EAOGe-C$_2$• bonds to the IR-C$_4$ structure instead of the intended IR-C$_2$, yielding a pendent C$_6$• intermediate that relaxes to the surface in a non-IR configuration. In the second, EAOGe-C$_2$• bonds to the IR-C$_2$ as intended, but the resulting pendent C$_4$• forms a C–C bond with the neighbouring IR-C$_4$ instead of relaxing to the IR configuration. Further work is required to distinguish these outcomes, to be able to utilize this structure for novel potential build targets, or to optimize conditions to avoid its formation in favor of improved IR-C$_4$ yield. All other off‑target outcomes were isolated, unrepeated features without an identifiable pathway, collectively accounting for 3\% of interaction outcomes (11 out of 351; 95\% confidence interval, 2-6\%).
\vspace{1em} 

\noindent These mechanosynthetic capabilities enable access to atomically precise patterns and structures not available through conventional surface chemistry. IR‑C$_2$ has previously been generated on silicon through electron‑induced dehydrogenation of adsorbed acetylene, but in that approach, positioning is governed by stochastic adsorption rather than deterministic placement\cite{MacLean2026}. IR‑C$_4$ is a previously unreported structure, demonstrating that mechanosynthesis can access novel covalently bonded species. Together, these results establish atomic-scale control over composition, bonding, and spatial placement, opening avenues to the assembly of bespoke, three-dimensional carbon architectures on silicon and providing a foundation for integration of mechanosynthetic APF with molecular‑scale electronics and silicon-based quantum devices.
\vspace{4em}

\noindent \textbf{Acknowledgements} 
\vspace{0.4em}

\noindent We thank the hardware team, including Byoung Choi, Sheldon Haird, William Cullen, and David Lister, for maintaining the laboratory infrastructure and for developing the custom hardware used for IM‑STM‑enabled mechanosynthesis. We also thank Eduardo Barrera Ramirez and Cristina Yu for SPC fabrication, Darian Blue for organizational support in preparing this manuscript, and Steven DeSmet for project and program management.
\vspace{3em}

\noindent \textbf{Funding} 
\vspace{0.4em}

\noindent This study was supported by the Government of Canada (Innovation, Science and Economic Development Canada) Strategic Innovation Fund (SIF, Agreement Number 813022) and by The Canadian Bank Note Company, Ltd. (CBN) under CBN Nano Technologies Inc. (CBNNT).  
\vspace{1em} 

\newpage
\noindent \textbf{Author contributions} 
\vspace{0.4em}

\noindent Conceptualization: DA, JB, M. Drew, M. Durand, RF, AH, MK, RM, TT, MT, DV, RY
\vspace{0.2em}

\noindent Methodology: DA, AA, BB, AB, KC, MC, CD, JF, AG, RG, SYG, AI, MK, RK, HM, CM, OM, SM, M. Marshall, M. Morin, JM, RP, SR, LS, MS, KSA, TT, MT, BT, RW, RY
\vspace{0.2em}

\noindent Software: DA, KC, MC, CD, RG, AI, MK, CM, OM, M. Marshall, JM, LS, TT, MT, RW, RY
\vspace{0.2em}

\noindent Resources: RA, MC, CD, M. Drew, TE, RG, AH, TH, AI, RK, TM, M. Morin, JM, HR, SR, KSA, BS, BT, JW, RW, RY
\vspace{0.2em}

\noindent Investigation: DA, AA, BB, AB, MC, CD, JF, AG, RG, SYG, AI, RK, HM, AM, CM, OM, M. Marshall, JM, RP, LS, MS, KSA, MT, BT, RW, RY
\vspace{0.2em}

\noindent Data curation: DA, MC, CD, RG, AI, RK, CM, M. Marshall, JM, LS, KSA, BT, RW, RY
\vspace{0.2em}

\noindent Formal analysis: DA, AA, BB, AB, MC, CD, JF, AG, RG, SYG, AI, MK, RK, HM, CM, OM, M. Marshall, M. Morin, JM, RP, SR, LS, MS, KSA, MT, BT, RW, RY
\vspace{0.2em}

\noindent Visualization: MC, MJ, CM, JM, SR, LS, RW
\vspace{0.2em}

\noindent Funding acquisition: JB, RF, RM
\vspace{0.2em}

\noindent Project administration: M. Durand, RG, AI, KSA, TT, RY
\vspace{0.2em}

\noindent Supervision: DA, MC, KSA, TT, MT, RY
\vspace{0.2em}

\noindent Writing – Original draft: DA, MC, JM, SR, LS, KSA, BT, RW, RY
\vspace{0.2em}

\noindent Writing – Review and editing: RA, DA, MC, M. Durand, AG, RG, AH, MJ, RK, OM, M. Morin, LS, MS, BT, RW, RY
\vspace{4em}

\bibliographystyle{unsrt}
\bibliography{references}

\noindent{\color{white}\NoHyper
\cite{Pitters2024}
\cite{Barrera2025}
\cite{McCallum2026}
\cite{Shen1997, Pitters2024}
\cite{Barrera2025}
\cite{Tersoff1985}
\cite{Ufimtsev2009, Titov2012, Song2015}
\cite{Becke1993}
\cite{Chai2008}
\cite{Hehre1972}
\cite{Wadt1985}
\cite{Lin2012}
\cite{Weigend2006}
\cite{Gaussian16}
\cite{HjorthLarsen2017}
\cite{Chai2008}
\cite{Ufimtsev2009, Titov2012, Song2015}
\cite{Wentzel1926, Kramers1926, Brillouin1926}
\cite{Henkelman2000}
\cite{HjorthLarsen2017}
\cite{Biedermann1999, Lastapis2005, Bellec2010}
\endNoHyper}

\end{document}